\documentstyle[seceq,epsf,wrapft]{ptptex}

\notypesetlogo  

\title{
\Large \bf Cosmological Simulation of \\ Gravitational Weak Lensing 
in Open Models}

\author{%
Kenji {\sc Tomita}
}

\inst{%
Yukawa Institute for Theoretical Physics, 
Kyoto University, Kyoto 606-8502
}

\abst{%
Cosmological simulation of gravitational weak lensing in open
models with $(\Omega_0, \lambda_0) = (0.2,0)$ and $(0.4,0)$
is treated using the direct integration method, which is independent of the
multi-lens-plane method. Statistical averages of optical quantities
(the convergence, shear, and amplification) are derived for ray
bundles passing through the clumpy universe, in which the initial
state has the CDM spectrum with the power $n = 1$, and all particles 
are regarded as equivalent (compact) lens objects with a galactic
size. The compact lens objects consist of galaxies and dark matter particles.
 The evolution of the spatial mass distribution is given by $N$-body 
simulations in a tree-code. This paper is an extension of our previous 
one in flat models.
The behaviors of optical quantities are shown in comparison with a
low-density flat model with  $(0.2,0.8)$.  
It is found as a result that (1) optical quantities in the open 
model increase with $\Omega_0$, and \ (2) they are larger than those in
the flat model with the same $\Omega_0$, and the difference is
comparable with that between two open models with $\Omega_0 = 0.2$ and
$0.4$.
The comparison of the optical quantities with the corresponding observed
ones will give us some information on the observational value of $\Omega_0$.
}

\begin{document}

\maketitle

\section{Introduction}

Gravitational weak lensing is one of the most useful tools for probing 
the inhomogeneous distribution of mass in the universe. The small
deformation in the shape of distant galaxies caused by gravitational tidal
force is directly connected with the total mass distribution in the
neighbourhood of light paths to galaxies. Recently theoretical and
observational studies on the deformed images of distant
galaxies\cite{rf:bland}\tocite{rf:swm} have shown that in a near future the
lensing information on the mass distribution brought with large telescopes
will be useful to put severe constraints on dark matter, dark halos 
and allowed values of cosmological parameters.  

In a previous paper\cite{rf:t} we treated the statistics of
gravitational weak lensing in flat cosmological models in the direct
integration method, in which the ray shooting was performed by
solving null-geodesic equations numerically in the intervals between 
an observer and sources. Here weak lensing does not mean the weakest
limit of lensing, but means that we treat only rays without caustics. 

In this paper we extend the treatment to the case of open models and
derive the average values of optical quantities on various angular
scales, in comparison with those in the flat case. The largest
difference of treatments in curved cases from the flat cases 
is that in curved cases we can construct no model universes covered 
everywhere with periodic boxes, and so, for the calculation of the
gravitational forces from particles, we must assume a set of periodic
boxes which are closely connected only along each light ray. 

In \S 2 we show treated model universes and the equations for ray shooting
in open models, and in \S 3 derive the statistical averages of optical 
quantities. Their behaviors are shown in connection with those of the
angular diameter distances. In \S 4 the comparison of average optical 
quantities with observed ones is discussed for the estimate of $\Omega_0$.
In Appendices A the derivation of null-geodesic equations
is shown in open models.

\section{Model universes and the ray shooting}

The background model universes are assumed to be spatially open or
have the negative curvature, and
the behavior of light rays passing through inhomogeneities in them is
 considered at the stage from an initial time $t_1$ (with redshift
$z_1 = 5$) to the present time $t_0$. In
the Newtonian approximation, the line-element is expressed as 
\begin{equation}
  \label{eq:ba1}
ds^2 = -(1+2\varphi/c^2) c^2 dt^2 
+ (1-2\varphi/c^2)a^2(t)(d\mib{x})^2/[1 + K {1 \over 4}(\mib{x})^2]^2,
\end{equation}
where $K$ is the signature of spatial curvature $(\pm 1, \ 0)$. In the 
following we take $K = -1$.
The normalized scale factor $S \equiv a(t)/a(t_0)$ satisfies
\begin{equation}
  \label{eq:ba2}
 \Bigl({dS \over d \tau} \Bigr)^2 = {1 \over S} \Bigl[\Omega_0 -
(\Omega_0+\lambda_0-1)S  + \lambda_0 S^3 \Bigr],
\end{equation}
where $\tau \equiv H_0 t$ and $a_0 (\equiv a(t_0))$
is specified by  a relation 
$(c H_0^{-1} /a_0)^2 = 1 - \Omega_0 - \lambda_0.$
 The gravitational potential $\varphi$ is described by the Poisson equation
\begin{eqnarray}
  \label{eq:ba3}
a^{-2} \Delta \varphi &=& [1 - {1 \over 4}(\mib{x})^2]^2  \Bigl[{\partial^2 
\varphi \over \partial \mib{x}^2} + {{1 \over 2}x^i \over  [1 - {1 \over 4}
(\mib{x})^2]^3} {\partial \varphi \over \partial x^i}\Bigr] \cr
&=& 4 \pi G \rho_B [\rho(\mib{x})/\rho_B -1],
\end{eqnarray}
where $\rho_B (=\rho_{B0}/S^3)$ is the background density and 
\begin{equation}
  \label{eq:ba4}
\rho_{B0} = {3 {H_0}^2 \Omega_0 \over 8\pi G} = 2.77 \times 10^{11}
\Omega_0 h^2 M_\odot \ {\rm Mpc}^{-3},
\end{equation}
where $H_0 = 100 h {\rm Mpc^{-1} \ km \ s^{-1}}$. 
In our treatment the inhomogeneities are locally periodic in the
sense that the physical situation at $\mib{x}$ is the same as that at
$\mib{x} + l \mib{n}$, where the components of $\mib{n} 
(=(n^1,n^2, n^3))$ are
integers. In an arbitrary periodic box with coordinate volume $l^3$, 
there are $N$ particles with the same mass $m$. It is assumed that the 
force at an arbitrary point is the sum of forces from $N$ particles in 
the box whose center is the point in question, and that the forces from
outside the box can be neglected. 

In this paper we take two open models (O1 model and O2 model) with
$(\Omega_0, \lambda_0) = (0.2, 0)$ and $(0.4, 0)$, respectively, and
a flat model (L model) with $(0.2, 0.8)$ for comparison. The 
present lengths of the boxes are
\begin{equation}
  \label{eq:ba5}
L_0 \equiv a(t_0) l = 50 h^{-1}, \ 39.7 h^{-1} {\rm Mpc}
\end{equation}
for $\Omega_0 = 0.2, 0.4$, respectively. The particle number is $32^3$ 
in both models, and so
\begin{equation}
  \label{eq:ba6}
m (= \rho_{B0} {L_0}^3/N) = 2.11 \times 10^{11} h^{-1} M_\odot.
\end{equation}

The distributions of particles in these models were derived by the numerical
$N$-body simulations using Suto's tree-code \cite{rf:su}
during the time interval $z
= 0$ and $z = z_1$, where $z_1 = 5.0$ for all models.
All particles in these models are regarded as 
equivalent (compact) lens objects with a galactic size, which consist of 
galaxies and dark matter.

For the compact lens objects we assume the physical softening radius 
$a(t) x_s = 20  h^{-1}  {\rm kpc}$. 

Light propagation is described by solving the null geodesic
equation with the null condition. Here let us use $T \equiv 
{1 \over 2}\ln
[a(t)/a(t_1)]$ as a time variable and $T_0 \equiv (T)_{t = t_0}$. Then
we have $dS = 2 \exp [2(T- T_0)] dT$, so that 
\begin{equation}
  \label{eq:ba7}
c dt = R\ c_R \ [\Omega_0 +(1 -\Omega_0 -\lambda_0) S + \lambda_0 
S^3]^{-1/2} {e}^{3 T} dT,
\end{equation}
where
\begin{equation}
  \label{eq:ba8}
R \equiv L_0/[(1+z_1) N^{1/3}]
\end{equation}
and
\begin{equation}
  \label{eq:ba9}
c_R \equiv 2(c/H_0)/[R(1+z_1)^{3/2}].
\end{equation}
The line-element is 
\begin{eqnarray}
  \label{eq:ba10}
ds^2/R^2 = -{c_R}^2  [\Omega_0 &+&(1 -\Omega_0 -\lambda_0) S + 
\lambda_0 S^3]^{-1} {e}^{6T} (1+\alpha \phi) dT^2 \cr 
&+& (1-\alpha \phi) {e}^{4T} d\mib{y}^2 /F(\mib{y})^2,
\end{eqnarray}
where \ $y^0 = T, \ y^i = a(t_1) x^i/R, \ \varphi = (G m/ R) \phi$, \ 
$R_0 \equiv R a_0/a_1 = (1+z_1) R$, 
\begin{equation}
  \label{eq:ba11}
\alpha \equiv {2Gm \over c^2 R} = {3 \over \pi} {\Omega_0 \over (c_R)^2},
\end{equation}
and
\begin{equation}
  \label{eq:ba11a}
F \equiv 1 -{1 \over 4}(R_0H_0/c)^2 (1 -\Omega_0 -\lambda_0) 
(\mib{y})^2.
\end{equation}
The equations for light rays to be solved are 
\begin{equation}
  \label{eq:ba12}
{dy^i \over dT} = c_R  {e}^T \tilde{K}^i,
\end{equation}
\begin{eqnarray}
  \label{eq:ba13}
{d\tilde{K}^i \over dT} &=& - [3 \lambda_0 e^{4(T-T_0)}+(1 -\Omega_0 
-\lambda_0)]  e^{2(T-T_0)} \tilde{K}^i /G(T)
  +\alpha {\partial \phi 
\over \partial T} \tilde{K}^i \cr
&-&\gamma {c_R}^{-1}  {e}^T \Bigl[{\partial \phi /
\partial y^i}/G(T) - 2 {\partial \phi \over
\partial y^j} \tilde{K}^j \tilde{K}^i\Bigl] \cr
&+& (R_0H_0/c)^2 (1 -\Omega_0 -\lambda_0) F^{-1} c_R e^T \Bigl[-y^j
\tilde{K}^j \tilde{K}^i \cr
&+& {1 \over 2}(1 +2\alpha \phi)/G(T) \Bigl] ,
\end{eqnarray}
where $\gamma \equiv \alpha (c_R)^2$ \ and
\begin{equation}
  \label{eq:ba13a}
G(T) = \Omega_0 +(1 -\Omega_0 
-\lambda_0) e^{2(T-T_0)}+ \lambda_0  e^{6(T-T_0)}.
\end{equation}
The null condition is 
\begin{equation}
  \label{eq:ba14}
\sum_i (\tilde{K}^i)^2 = 1 + 2\alpha \phi.
\end{equation}
The derivation of these equations is given in Appendix A.

The potential $\phi$ is given as a solution of the Poisson equation.
Because the ratio of the second term to the first term in the
right-hand side of Eq. (\ref{eq:ba3}) is $(R_0H_0/c)^2 (\mib{y})^2
[\Delta y/\vert \mib{y}\vert] << 1$ for $z \leq z_1$, the Poisson
equation in a box can be approximately expressed as
\begin{equation}
  \label{eq:ba15}
F(\mib{y}_c)^2 \ {\partial^2 \phi \over \partial \mib{y}^2} =
4 \pi G \rho_B [\rho(\mib{x})/\rho_B -1],
\end{equation}
where we used that $F(\mib{y}) \simeq 1$ for $y \sim$ the box size and
can be approximately replaced by the central value $F(\mib{y}_c)$ 
for $y >>$ the box size and the suffix $c$ denotes the central value 
in the box. 
For point sources with $ \rho  = m \sum_n \delta(a R (\mib{y} 
-{\mib{y}_n}))$ \ ($n$ is the particle number), we have 
\ $\phi = \phi_1 + \phi_2$, \ where   
\begin{equation}
  \label{eq:ba16}
\phi_1 = - F(\mib{y}_c) e^{-2T} \sum_n {1 \over \vert \mib{y} 
-\mib{y}_n\vert},
\end{equation}
and $\phi_2$ displays the contribution from the homogeneous background
density. Here let us use for $\mib{y}$ another coordinates $\bar{\mib{y}}$
expressing the space in a locally flat way, where the two coordinates
are connected as 
\begin{equation}
  \label{eq:ba17}
\bar{\mib{y}} = \int^{\mib{y}}_{\bf 0} d\mib{y}/F(\mib{y}),
\end{equation}
and the lengths between two points in boxes in two coordinates are 
approximately related as
\begin{equation}
  \label{eq:ba18}
\Delta \bar{\mib{y}} = \Delta \mib{y}/F(\mib{y}_c).
\end{equation}
Then $\phi_1$ is expressed in terms of $\bar{\mib{y}}$ in the usual
manner as
\begin{equation}
  \label{eq:ba19}
\phi_1 = - e^{-2T} \sum_n {1 \over \vert \bar{\mib{y}} 
-\bar{\mib{y}}_n\vert}.
\end{equation}
Corresponding forces are expressed as $f_i \equiv \partial
\phi_1/ \partial y^i = [\partial \phi_1/ \partial {\bar y}^i]/F(\mib{y}_c)$.
It should be noted that the contribution of $\partial \phi/ \partial T$ is 
negligibly small, compared with that of $f_i$.

In flat models the universe is everywhere covered with periodic boxes
continuously connected as in Fig. 1. In open models it cannot be covered
in a similar way, but we can consider only a set of local periodic boxes
connected along each light ray, as in Fig. 2. In these boxes we can 
describe the evolution in the distribution of particles in terms of 
local flat coordinates $\bar{\mib{y}}$, because the size of boxes is 
much smaller than the curvature radius. 

\begin{figure}
\epsfxsize=5.5cm
\centerline{\epsfbox{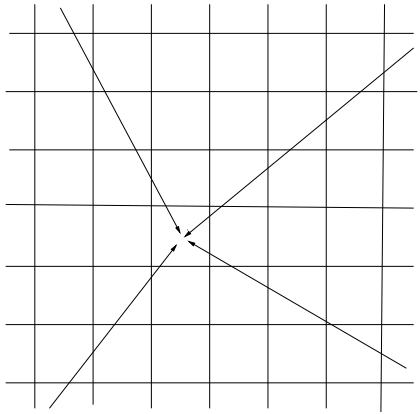}}
\caption{Light rays and periodic boxes in the flat space.}
\label{fig:1}
\end{figure}

\begin{figure}
\epsfxsize=5.5cm
\centerline{\epsfbox{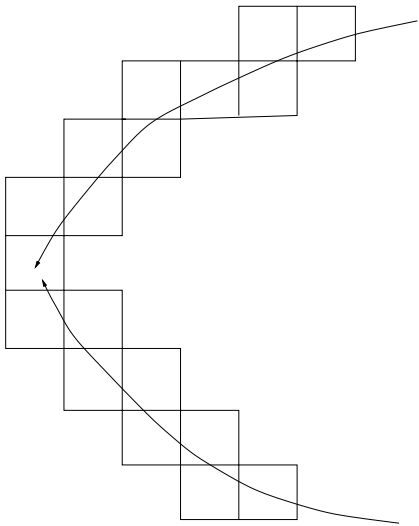}}
\caption{Light rays and local periodic boxes shown schematically in a
curved space.}
\label{fig:2}
\end{figure}

The time evolution in the distribution of particles was derived by
performing the $N$-body simulation in the tree-code provided by Suto.
The initial particle distributions were derived using Bertschinger's
software {\it COSMICS}\cite{rf:bert}
 under the condition that their perturbations are given
as random fields with the spectrum of cold dark matter, their power
$n$ is 1, and their normalization is specified as the dispersion 
$\sigma_8 \ = \ 0.94$ with the Hubble constant $h = 0.7$.

For the integration of the above null-geodesic equations, we calculate 
the potential at a finite number of points on the ray which are 
given at each time step ($\Delta T$). Then particles near one of the 
points on the ray have a stronger
influence upon the potential than particles far from any points. 
To avoid this unbalance in the calculation of the potential, we take
an average of the potential $\phi_1$ by integrating it analytically 
over the interval between one of the points and the next point. 
The expression for an averaged potential $\bar \phi$ was given in the 
previous paper. Moreover, to take into account the finite particle
size as galaxies or clouds,
we modify the above potential for point sources using the 
softening radii $y_s = a(t_1) x_s/R$. The modified potential is
produced by replacing $(y-y_n)^2$ to $(y-y_s)^2 + (y_s)^2$ in the  
potential for point sources.  

The initial values of $\tilde{K}^i$ are given so as to satisfy
Eq. (\ref{eq:ba14}). The integration of Eqs. (\ref{eq:ba12}) and
(\ref{eq:ba13}) with the modified potential was performed using the Adams 
method as in our previous papers. As the time step we assumed $\Delta
T = [\ln 6/2]/N_s$ with $/N_s = 3000$ (in most cases) $ - 10000$. 

\section{Statistical behavior of optical quantities}

We treat the deformation of ray bundles over the interval from $z = 0$ 
to $z = 5$ measured by an observer in a periodic box. In the same way 
as in the previous paper\cite{rf:t}, we consider 
here the ray bundles reaching the observer in a regular form such that 
the rays are put in the same separation angle $\theta$, and
calculate the change in the angular positions of the rays increasing
with redshift in the past direction. From this change we find the
behavior of optical quantities.

Basic ray bundles consist of  5 $\times$ 5  rays which are put in the square
form with the same
separation angle $\theta = 2 - 360$ arcsec. Many bundles coming 
from all directions in
the sky are considered. Here we take 200 bundles coming from randomly
chosen directions for each separation angle. In order to express 
angular positions of rays, we use two orthogonal vectors $e^i_{(1)}$
and $e^i_{(2)}$ in the plane perpendicular to the first background ray 
vector $(\tilde{K}^i)_B$.  
Then the angular coordinates $[X(m,n), Y(m,n)]$ of 25 rays relative to 
the first ray with $(m,n) = (1,1)$ at any epoch are defined by
\begin{eqnarray}
  \label{eq:st2}
X(m,n) &=& \sum_i [y^i(m,n) -  y^i(1,1)]\ e^i_{(1)}/y_B (1,1) +X(1,1),\cr
Y(m,n) &=& \sum_i [y^i(m,n) -  y^i(1,1)]\ e^i_{(2)}/y_B (1,1) +Y(1,1),
\end{eqnarray}
where $y^i = y^i_B + \delta y^i$ and $y_B = [\sum_i (y^i_B)^2]$.
Since all angular intervals of rays at the observer's point are the same ($=
\theta$), differentiation of angular coordinates of the
rays at any epoch with respect to those at observer's points is given
by the following differences:
\begin{eqnarray}
  \label{eq:st3}
A_{11}(m,n) &=& [X(m+1,n) - X(m,n)]/\theta, \cr
A_{12}(m,n) &=& [X(m,n+1) - X(m,n)]/\theta, \cr
A_{21}(m,n) &=& [Y(m+1,n) - Y(m,n)]/\theta, \cr
A_{22}(m,n) &=& [Y(m,n+1) - Y(m,n)]/\theta, 
\end{eqnarray}
where $m$ and $n$ run from 1 to 5. From the matrix $A_{ij} (m,n)$ we
derive the optical quantities in the standard manner,\cite{rf:sef}
as the convergence ($\kappa (m,n)$),
the shear ($\gamma_i (m,n), \ i=1,2$), and the amplification ($\mu
(m,n)$) defined by
\begin{eqnarray}
  \label{eq:st4}
 \kappa (m,n) &=& 1 - {\rm tr}(m,n)/2, \quad \gamma_1 (m,n) = [A_{22}(m,n) - 
A_{11}(m,n)]/2, \cr
\gamma_2 (m,n) &=& -[A_{12}(m,n) + A_{21}(m,n)]/2, \cr
\gamma^2 &\equiv& (\gamma_1)^2 + (\gamma_2)^2 = [{\rm tr}(m,n)]^2 - 
\det(A_{ij}(m,n)), \cr
\mu (m,n) &=& 1/\det(A_{ij}(m,n)),
\end{eqnarray}
The average optical quantities in each bundle are defined as the
averages of optical quantities for all rays in the bundle as follows:
\begin{equation}
  \label{eq:st5}
\bar{\kappa} = \Bigl[\sum_m \sum_n \kappa (m,n)\Bigr]/4^2, \quad
\bar{\kappa^2} = \Bigl[\sum_m \sum_n (\kappa (m,n))^2\Big]/4^2,
\end{equation}
and so on. In this averaging process the contributions from smaller 
scales can be cancelled and smoothed-out. The above optical quantities 
at the separation angle $\theta$ are accordingly derived in the 
coarse-graining on this smoothing scale. 

For the present statistical analysis we excluded the caustic cases
and considered only the cases of weak lensing in the sense of no 
caustics.
The averaging for all non-caustic ray bundles is denoted using $< >$
as $<\kappa^2>, \ <\gamma^2>$  and  $<(\mu -1)^2>$. Because
$\kappa(m,n), \ \gamma_i (m,n)$ and $\mu (m,n) -1$ take positive and
negative values with almost equal frequency, $<\kappa>, \ <\gamma_i>$ 
and $<\mu> -1$  are small. 

In Figs. 3 and 4, we show the behavior of $<\kappa^2>$ and 
$<\gamma^2>$ for various separation angles
$\theta = 2$ arcsec  $- \ 360$ arcsec (= $6$ arcmin). 
It is found that  $<\kappa^2>^{1/2}$ and $<\gamma^2>^{1/2}$ for 
$\Omega_0 = 0.4$ \ (model O2) are larger than those for $\Omega_0 = 
0.2$ \ (model O1) at all separation angles $\theta = 2 - 360$ arcsec 
and the difference $\Delta$ is about $0.005 - 0.01$ at both periods 
$z = 1$ and $2$, and that for $\Omega_0 = 0.2$ those in the open model 
(O1) are larger than those in the flat model (L) and their difference 
$\Delta'$ is comparable with the above $\Delta$. The amount of optical 
quantities for $\theta \sim 360$ arcsec is consistent with the results 
of Bernardeau et al. (cf. their Figs. 3 and 4)\cite{rf:bern} and 
Nakamura\cite{rf:nakam} in their treatments. The differences of 
quantities between various cosmological models are closely connected 
with the differences of angular diameter distances in their models, 
because for the same separation angle larger angular diameter
distances correspond to longer inhomogeneities with smaller amounts
and so smaller deformations are brought to rays passing through them. 
Here let us examine the behavior of angular diameter distances to
explain this situation. In weak lensing the Friedmann angular 
diameter distances can be used approximately to most rays, and their 
behaviors are shown in Fig.5 for models with
$(\Omega_0, \lambda_0) = (1, 0), (0.4, 0), (0.2, 0)$ and $(0.2, 0.8)$.
The basic equations for deriving this distances can be seen in a standard
text.\cite{rf:sef} 
From this figure we find that these models stand in inverse order of length
of the distances, and so that in model L the distance is larger for
equal redshifts and rays with the same separation angles have smaller 
deformations than in model O1.  

In Tables I, II and III, we show for models O1, O2 and L the 
numerical values of $<\kappa>,\ 
<\kappa^2>^{1/2}, \ <\mu>, \ <(\mu -1)^2>^{1/2}, \  <\gamma_1>$ and 
$<\gamma^2>^{1/2}$ \ at epochs $z = 1,2, .., 5$ for $\theta = 2$ 
arcsec. 

\begin{figure}
\epsfxsize=7.5cm
\centerline{\epsfbox{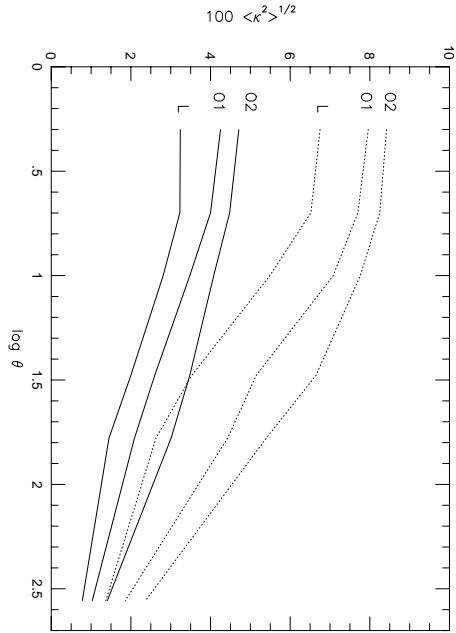}}
\caption{The angular dependence of $<\kappa^2>^{1/2}$. 
Solid and dotted lines denote behaviors for $z = 1$ and $2$, respectively. 
The separation angle $\theta$ is in the unit of arcsec.}
\label{fig:3}
\end{figure}

\begin{figure}
\epsfxsize=7.5cm
\centerline{\epsfbox{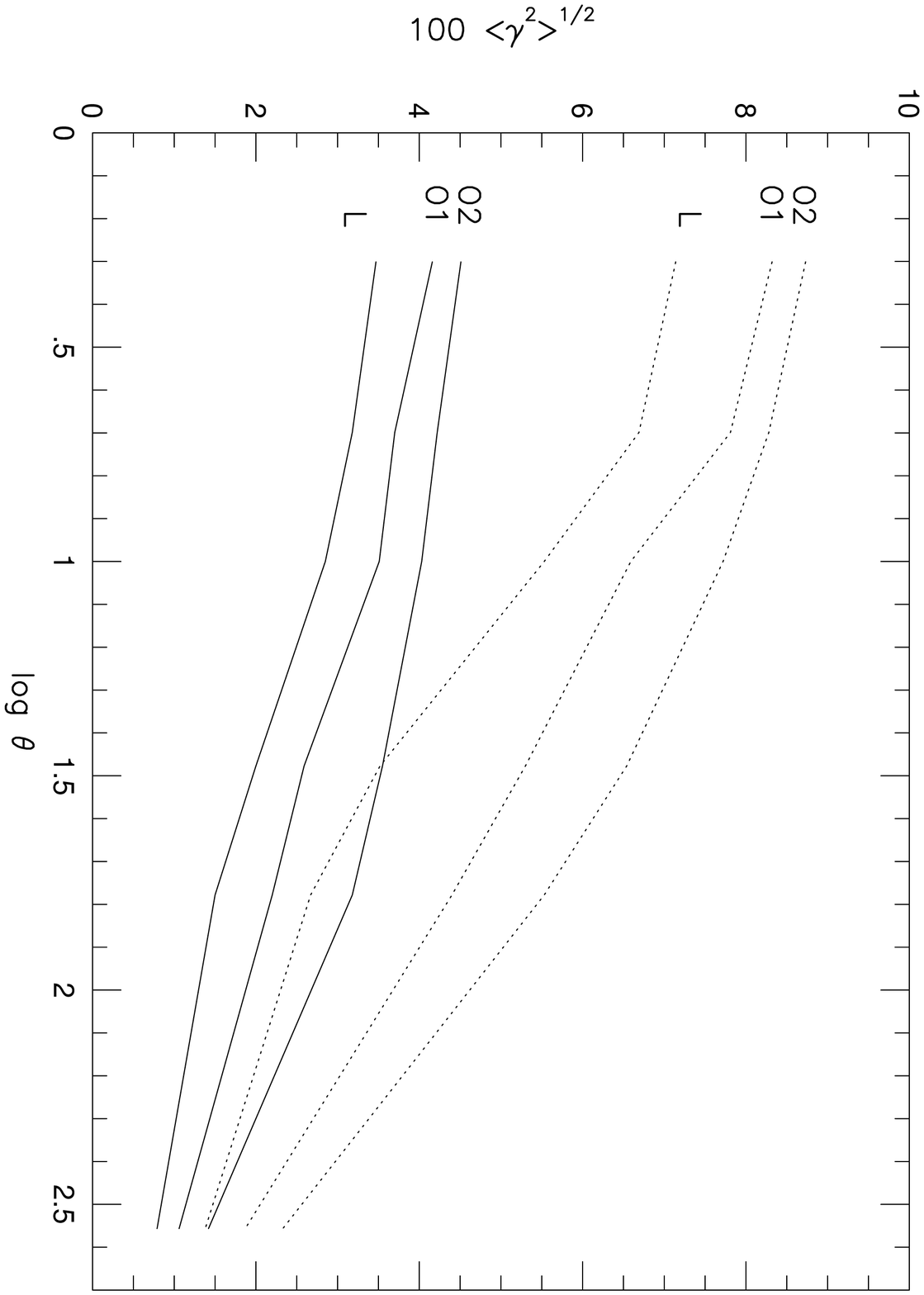}}
\caption{The angular dependence of $<\gamma^2>^{1/2}$.
Solid and dotted lines denote behavior for $z = 1$ and $2$, respectively. 
The separation angle $\theta$ is in the unit of arcsec.}
\label{fig:4}
\end{figure}

{
\begin{wraptable}{l}{\halftext}
\caption{Optical quantities in model O1.} 
\label{table:1}
\begin{center}
\begin{tabular}{cccccrc} \hline \hline
$z$&$<\kappa>$ & $<\kappa^2>^{1/2}$  & $<\mu>$ & $<(1-\mu)^2>^{1/2}$
 & $<\gamma_1>$ & $<\gamma^2>^{1/2}$
 \\ \hline 
1&   $-0.0002$ &  $0.0425$ &  1.0221 &  0.1219 & $-0.0028$ &  0.0426\\
2&   $-0.0054$ &  $0.0796$ &  0.9961 &  0.2210 & $-0.0041$ &  0.0832\\
3&   $-0.0069$ &  $0.1157$ &  1.0368 &  0.6234 & $ 0.0057$ &  0.1225\\
4&   $-0.0081$ &  $0.1519$ &  1.0790 &  0.5638 & $ 0.0078$ &  0.1602\\
5&   $-0.0064$ &  $0.1872$ &  1.1627 &  1.1880 & $ 0.0120$ &  0.1980\\ \hline
\end{tabular}
\end{center}
\bigskip

\caption{Optical quantities in model O2. } 
\label{table:2}
\begin{center}
\begin{tabular}{cccccrc} \hline \hline
$z$&$<\kappa>$ & $<\kappa^2>^{1/2}$  & $<\mu>$ & $<(1-\mu)^2>^{1/2}$
 & $<\gamma_1>$ & $<\gamma^2>^{1/2}$
 \\ \hline 
1&  $0.0008$ &  0.0471 &  1.0126 &  0.1305 & -0.0011 &  0.0451\\
2&  $0.0071$ &  0.0842 &  1.0517 &  0.2755 & -0.0001 &  0.0873\\
3&  $0.0111$ &  0.1103 &  1.0980 &  0.4956 & -0.0022 &  0.1255\\
4&  $0.0136$ &  0.1380 &  1.1602 &  1.0742 & -0.0019 &  0.1613\\
5&  $0.0147$ &  0.1680 &  1.1926 &  1.5755 & -0.0028 &  0.1958\\ \hline
\end{tabular}
\end{center}

\bigskip

\caption{Optical quantities in model L. } 
\label{table:3}
\begin{center}
\begin{tabular}{cccccrc} \hline \hline
$z$&$<\kappa>$ & $<\kappa^2>^{1/2}$  & $<\mu>$ & $<(1-\mu)^2>^{1/2}$
 & $<\gamma_1>$ & $<\gamma^2>^{1/2}$
 \\ \hline 

1&   $-0.0011$ &  0.0324 &  1.0025 &  0.0705 &  0.0008 &  0.0347\\
2&   $-0.0036$ &  0.0675 &  1.0134 &  0.1572 &  0.0018 &  0.0714\\
3&   $-0.0045$ &  0.0974 &  1.0345 &  0.2407 &  0.0020 &  0.1002\\
4&   $-0.0065$ &  0.1216 &  1.0560 &  0.3209 &  0.0023 &  0.1214\\
5&   $-0.0070$ &  0.1405 &  1.0805 &  0.4000 &  0.0030 &  0.1372\\ \hline
\end{tabular}
\end{center}
\end{wraptable}
}

\section{Concluding remarks}
In this paper we assumed that all low-density models (O1, O2 and L)
contain only compact lens objects with the same mass and radius, and 
compared their optical quantities. If we assume that particles in
model O2 consist of compact lens objects and clouds with same mass but 
larger radius and that the number of compact lens objects is equal to that
in models O1 and L, the values of optical quantities in model O2 will 
decrease as the radius of clouds increase, just as S(b) and S(c) in 
the Einstein-de Sitter model S, which were shown in the previous 
paper.\cite{rf:t}

For the statistical analysis we used 200 ray bundles reaching an
observer in a single inhomogeneous model universe. This number of ray
bundles may be too small to cover the influences from complicated
inhomogeneities in all directions. For getting more robust statistical 
results it may be necessary to use more ray bundles and more model
universes produced with random numbers.

Here we touch recent observations of cosmological shear due to weak
lensing and their relation to our results.  Fort et al.\cite{rf:fort}
 attempted the 
measurements of a coherent shear from foreground mass condensations in 
the fields of several luminous radio sources.  Schneider et
al. \cite{rf:sch}
determined the shear in the field ($2$ min $\times 2$ min)
containing a radio source PKS1508-05 with $z = 1.2$, and their result
is that the shear is about 0.03 for the angular scale 1 min.
If we interpret this value as an upper limit of the average shear in
the general field due to weak lensing, we find from our result in Fig.
4 that $\Omega_0$ is $\sim 0.4$ at large. For its more confident
estimate, more observational data of coherent shears in various fields 
are required.

\begin{figure}
\epsfxsize=7.5cm
\centerline{\epsfbox{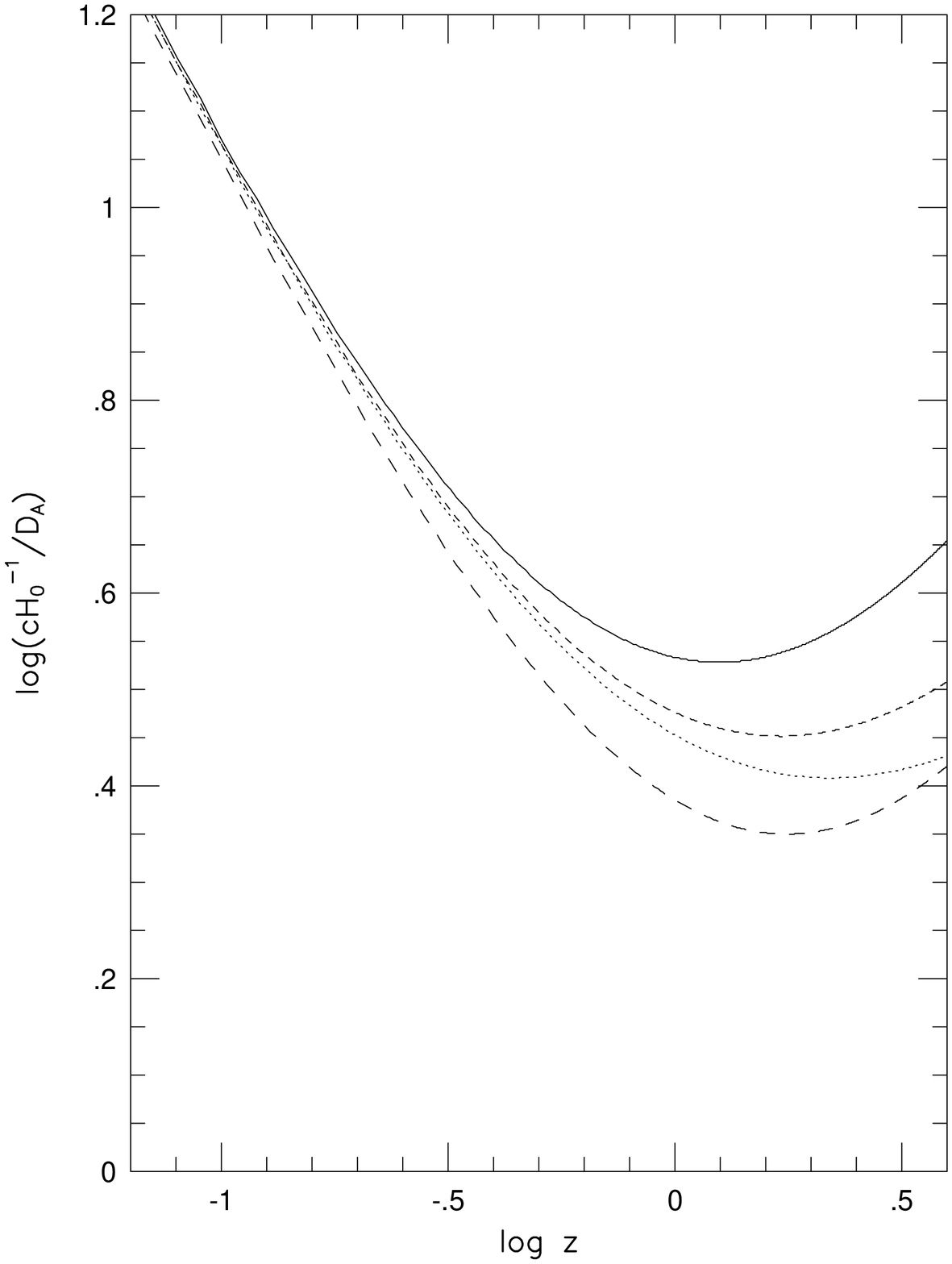}}
   \caption{The $z$ dependence of the angular diameter distance
$D_{\rm A}$ for $\alpha =
1$. Models S(1,0), O1(0.2,0), O2(0.4,0) and L(0.2,0.8) are denoted 
by solid, dotted, short dashed, long dashed lines, respectively. }
\label{fig:5}
\end{figure}

\section*{Acknowledgments}
The author would like to thank Y.~Suto, M.~Itoh, and K.~Yoshikawa for
helpful discussions about $N$-body simulations.
Numerical calculations were performed on the YITP computer system.

\appendix
\section{Basic Equations of Null Rays}

Corresponding to the line-element (\ref{eq:ba10}), we consider first the
wave-vectors $K^\mu \equiv (K^0, K^i) = dy^\mu/ dv$, where $v$ is an
affine parameter. The null condition is 
\begin{eqnarray}
  \label{eq:A1}
{\cal L} \equiv &-& [\Omega_0+(1 -\Omega_0 -\lambda_0) S +
\lambda_0 S^3]^{-1} e^{6T} (1+\alpha \phi) (K^0)^2 \cr
&+& {c_R}^{-2}(1-\alpha \phi)e^{4T}\sum_i (K^i)^2 
/F(\mib{y})^2 = 0,
\end{eqnarray}
where $F$ is defined in Eq.(\ref{eq:ba11a}).
From the equation 
\begin{equation}
  \label{eq:A2}
{d \over dv}{\partial {\cal L} \over \partial K^{\mu}} = {\partial 
{\cal L} \over \partial y^{\mu}},
\end{equation}
we obtain 
\begin{eqnarray}
  \label{eq:A3}
{d K^0 \over dv}&+& {3\Omega_0 +2(1 -\Omega_0-\lambda_0) e^{2(T-T_0)}
\over G(T)} (K^0)^2 + {2 G(T)
\over {c_R}^2 {e}^{2T} F^2 }\sum_i (K^i)^2  \cr
&+& \alpha \Bigl\{{1 \over 2} 
{\partial \phi \over \partial T}\Bigl[(K^0)^2 
- G(T) {c_R}^{-2}{e}^{-2T}\sum_i (K^i)^2/F^2 \Bigr]
 + {\partial \phi \over \partial y^i} K^i K^0  \cr
&-& 2 {c_R}^{-2}{e}^{-2T} \phi G(T) \sum_i (K^i)^2/F^2 \Bigr\} = 0,
\end{eqnarray}
\begin{eqnarray}
  \label{eq:A4}
{d K^i \over dv}&+& \Bigl(4 - \alpha {\partial \phi \over \partial T}
\Bigr) K^0 K^i - \alpha {\partial \phi \over \partial y^j}K^i K^j \cr
&+& {1 \over 2} \alpha {\partial \phi \over \partial y^i} \Bigl[{c_R}^2 
e^{2T} G^{-1}(T) (K^0)^2 + \sum_l (K^l)^2/F^2 \Bigr] \cr
&+& (R_0H_0/c)^2 (1 -\Omega_0 -\lambda_0) F^{-1}  \Bigl[\sum_l y^l K^l
K^i
 - {1 \over 2}\sum_l (K^l)^2 y^i \Bigr] = 0, \qquad
\end{eqnarray}
where $G(T)$ is defined by Eq. (\ref{eq:ba13a}).

Transforming $v$ to $T$, these equations lead to 
\begin{equation}
  \label{eq:A5}
{d K^0 \over dT} + \Bigl[ {3\Omega_0 + 2 (1 -\Omega_0 
-\lambda_0) e^{2(T-T_0)}\over  G(T)} +2\Bigr] K^0 +
\alpha {\partial \phi \over \partial y^i} K^i = 0,
\end{equation}
\begin{eqnarray}
  \label{eq:A6}
{d K^i \over dT} &+& \Bigl(4 - \alpha {\partial \phi \over \partial T}
\Bigr) K^i - \alpha {\partial \phi \over \partial y^j}{ K^j K^i \over
K^0} + {e}^{2T}(c_R)^2 G^{-1}(T){\alpha \partial \phi \over \partial 
y^i}\cr
&+& (R_0H_0/c)^2 (1 -\Omega_0 -\lambda_0)F^{-1} \Bigl[{\sum_j y^j K^j
K^i /K^0} \cr
&-&{1 \over 2}  {e}^{2T}(c_R)^2 G^{-1}(T)(1 + 2 \alpha \phi) y^i K^0
\Bigr]  
= 0.
\end{eqnarray}

Moreover, if we use for $K^i$ 
\begin{equation}
  \label{eq:A7}
\tilde{K}^i \equiv {c_R}^{-1} {e}^{-T} K^i/K^0,
\end{equation}
we obtain Eqs. (\ref{eq:ba12}) and (\ref{eq:ba13}).

\end{document}